\documentclass[pdflatex,sn-basic]{sn-jnl}

\usepackage{graphicx}%
\usepackage{multirow}%
\usepackage{amsmath,amssymb,amsfonts}%
\usepackage{amsthm}%
\usepackage{mathrsfs}%
\usepackage[title]{appendix}%
\usepackage{xcolor}%
\usepackage{textcomp}%
\usepackage{manyfoot}%
\usepackage{booktabs}%
\usepackage{algorithm}%
\usepackage{braket}
\usepackage{algorithmicx}%
\usepackage{algpseudocode}%
\usepackage{listings}%
\usepackage{url}
\usepackage{ulem}  

\theoremstyle{thmstyleone}%
\theoremstyle{thmstyletwo}%

\theoremstyle{thmstylethree}%

\begin{document}

\title[Article Title]{Machine Learning for Arbitrary Single-Qubit Rotations on an Embedded Device}


\author[1]{\fnm{Madhav Narayan} \sur{Bhat}}
\equalcont{These authors contributed equally to this work.}
\author[2,3]{\fnm{Marco} \sur{Russo}}
\equalcont{These authors contributed equally to this work.}
\author[1]{\fnm{Luca P.} \sur{Carloni}}
\author[3]{\fnm{Giuseppe} \sur{Di Guglielmo}}
\author[3]{\fnm{Farah} \sur{Fahim}}
\author[3]{\fnm{Andy C. Y.} \sur{Li}}
\author[3]{\fnm{Gabriel N.} \sur{Perdue}}
\affil[1]{\orgname{Columbia University}, \orgaddress{\city{New York}, \state{NY}, \country{USA}}}
\affil[2]{\orgname{Politecnico di Torino}, \orgaddress{\city{Turin}, \country{Italy}}}
\affil[3]{\orgname{Fermi National Accelerator Laboratory}, \orgaddress{\city{Batavia}, \state{IL}, \country{USA}}}


\abstract{Here we present a technique for using machine learning (ML) for single-qubit gate synthesis on field programmable logic for a superconducting transmon-based quantum computer based on simulated studies.
Our approach is multi-stage.
We first ``bootstrap'' a model based on simulation with access to the full statevector for measuring gate fidelity.
We next present an algorithm, named adapted randomized benchmarking (ARB), for fine-tuning the gate on hardware based on measurements of the devices.
We also present techniques for deploying the model on programmable devices with care to reduce the required resources.
While the techniques here are applied to a transmon-based computer, many of them are portable to other architectures.}

\keywords{Quantum Computing, Quantum Hardware, Quantum Control, Superconducting Qubits}



\maketitle



\section{Introduction}

%
Quantum computers may offer exponentially more efficient implementations of some algorithms than classical computers~\cite{Maslov2019}.
One of the quantum computers' most promising technological implementations is based on superconducting transmon qubits~\cite{koch2007_transmon,transmonref}.
In particular, superconducting transmon-based quantum computers offer one of the more promising paths to fault-tolerant quantum computing \cite{PhysRevA.86.032324,10.1063/5.0082975,ang2022architectures}.
Nevertheless, control of quantum computers based on this technology is challenging because qubits must be chilled to millikelvin temperatures in dilution refrigerators.
The communication latency and bandwidth between qubits and classical control systems ``in the warm'' is highly constrained.
In this simulation-based study, we investigate a machine learning (ML) approach to qubit control that will be implemented on field-programmable logic that runs in the refrigerator proximate to the the qubit.
We present the algorithm translation, design-space exploration, and resource requirements; we also consider algorithm training requirements in the presence of unstable noise sources.

Field-programmable gate arrays (FPGAs) and embedded FPGAs (eFPGAs) are two technologies in the field of field-programmable logic. FPGAs are standalone chips that offer programmability, enabling reconfiguration of the hardware architecture after manufacturing for various applications, including ML. This flexibility makes FPGAs suitable for cases where updating hardware logic post-deployment is necessary. On the other hand, eFPGAs are integrated into System-on-Chips (SoCs) or application-specific integrated circuits (ASICs), providing a reconfigurable logic block within a larger device. This integration allows for the addition of programmability to static designs, matching the performance benefits of ASICs with the flexibility of programmability. eFPGAs can reduce power consumption and increase system performance by allowing for the hardware acceleration of ML tasks directly within the SoC, enhancing the functionality and efficiency of integrated circuits in quantum-control applications.

Transmon qubits are controlled with electromagnetic pulses, i.e., microwaves. Their amplitude, phase, and shape parameters (that we will call coefficients from now on to differentiate them from the neural network parameters) determine the resulting quantum state, which is often described by rotations of the state vector over the Bloch sphere.
We call these operations \emph{quantum gates}. The simplest rotation of an angle $\theta$ about the $X$-axis is denoted as an operation (or gate) $R_x(\theta)$.
Any arbitrary rotation can be achieved as the sequence of $Z$, $X$, and $Z$ rotations, so only $R_z\left(\phi\right)$ and $R_x\left(\theta\right)$ gates are necessary. 
Furthermore, since for transmon devices $R_z$ gates may be obtained virtually just by varying the phase of the pulses~\cite{virtualzgate}, we may focus on implementing only $R_x$ gates.


Quantum gates may be thought of as analog operations.
For example, the $R_x\left(\theta\right)$ gate is parameterized by a real number.
Measurements of quantum bits yield binary results in the chosen basis, but accurate quantum computation requires that the executed rotation closely matches the intended rotation.
We use fidelity, which measures the overlap between state vectors in Hilbert space, as the primary metric for gate accuracy.
Our problem is to find pulse coefficients that engineer operations with the highest possible fidelity.

On a side note, any Bloch sphere rotation may furtherly be decomposed in a sequence of $R_z, \sqrt{X}, \sqrt{X}^{-1}$ gates.
Since on transmon platforms the $R_z$ may be achieved ``virtually'' by varying the phase of the pulses, another strategy is to focus on only $2$ angles for the $R_x$ gate ($-\frac{\pi}{2}, +\frac{\pi}{2}$).
This simplifies the calibration costs in exchange for longer gate sequences for achieving a single rotation (two $R_x$ gates are required instead of one).
This is the benchmark strategy our technique must ultimately be compared to for the problem of qubit rotation.
Our technique may offer advantages in execution time or overall fidelity.
Additionally, our technique may find application in other control tasks, such as realizing high-fidelity parameterized gates for superconducting cavities \cite{Kudra2022quditcontrol,You2024quditcontrol}, generating angle-robust two-qubit entangling gates for trapped-ion qubits \cite{Jia2023trappedionscontrol} and optimizing pulse parameters to robustly drive neutral-atom qubits \cite{russo2023}.

Tools such as \texttt{Juqbox}~\cite{petersson2022optimal}, \texttt{Qiskit Pulse}~\cite{qiskitpulse}, \texttt{QuTiP}~\cite{qutip1,qutip2}, and \texttt{Cirq}~\cite{cirq_developers_2023_10247207} allow users to use simulation to study the effect of a pulse on a quantum device.
For the $R_x\left(\theta\right)$ gate, simulating multiple angles produces a corresponding set of coefficients.
These solutions traditionally run on workstations in the warm, i.e., outside of the refrigerator; they are viable for real-time control but not practically feasible as the number of qubits keeps scaling and the communication becomes a bottleneck.
Transferring those frameworks and their traditional hardware in the refrigerator is unfeasible because of the reduced space available and limited power budget to maintain the qubits at the operational temperature.
%
Another option would be to use simpler implementations based on lookup tables. However, this is also not feasible because interpolation between table entries is a low-accuracy strategy, and mitigation requires constructing increasingly larger tables with more complex operations.
Additionally, these tables are not flexible when noise drifts on the hardware.

This work explores a different approach: using a neural network to infer the proper pulse and running it on programmable devices, such as eFPGAs, in the refrigerator.
We trained our neural network on a large set of angles for an $R_x\left(\theta\right)$ gate, and the corresponding pulse coefficients (describing a B-spline) found by \texttt{Juqbox}.
A preliminary work~\cite{Xu_2022} explored this approach using the Mean-Squared Error (MSE) between the network-inferred coefficients and those found by \texttt{Juqbox} for training and testing.
%
However, that approach relied on utilizing quantum state vector information which may not be observed directly on real hardware.
Furthermore, it did not address the fact that superconducting qubits present some characteristics that are not stable. 
For example, the qubit anharmonicity (which describes the frequency spacing of the energy levels) varies with temperature cycles.
These characteristics change if a superconducting computer is warmed up and then cooled down again.
Because these characteristic values may not be known precisely, it is essential to have a method for estimating fidelity based on measurements taken on the hardware directly.
The method presented in this work proposes and implements an extension of Randomized Benchmarking (RB)  \cite{Wallman_2014} for statistically estimating the fidelity of gates.

\subsection{Prior work}
\label{sec:priorwork}
A number of papers can be found in the literature proposing methods to calibrate quantum gates. For instance, \cite{Rodionov_2014} and \cite{abughanem2024quantumprocesstomographyuniversal} use variations of Quantum Process Tomography (QPT); however, the computational complexity of QPT makes it unusable for a large set of gates to be optimized. In \cite{Li_2023}, instead, the system is modeled with a Hamiltonian and the gates are numerically optimized, performing a simulation of the system. Such a method is certainly useful as a starting point, but cannot be implemented on its own on a real device, as the actual system is inherently more complicated than the model and there is no proposed method to adjust to the real qubits. Our method starts in a similar way using Juqbox for optimizing the gates during a numerical simulation of the system, but that is only used as a warm-up; the parameters are then fine-tuned considering physical variations that model the differences that occur at the passage from the model to the real system. Finally, there are also papers that apply RB to non-Clifford gates, such as \cite{Onorati_2019}, but they do not provide any statistical backing to the fidelity estimates.

\subsection{Our contributions}
\label{sec:contributions}

This paper builds on previous work by some of the authors \cite{Xu_2022}.
Our primary extensions here are to adapt the algorithm to use a cost function based on measurements from a real quantum computer.
While we use simulated data in this work, we hide unobservable ``truth values'' (i.e., the wavefunction) from the simulation and only use the results of projective measurements of quantum states, as with real quantum hardware.
We develop a benchmark for assessing the fidelity of our gate synthesis algorithm and estimate the training costs in shots on a quantum computer.

\subsection{Organization of this paper}
\label{sec:orgofpaper}

In Section \ref{sec:datagen} we discuss the key features of our data generation process.
The entire study presented here uses simulation of qubits and not real devices.
Section \ref{sec:pretrain} discusses the first stage of training (referred to in this work as a ``bootstrap''), in which we utilize the simulation to provide an exact value for the state fidelity.
At this stage of training our loss function is defined by the Mean-Squared Error (MSE) between the predicted pulse and pulse computed when the data was originally generated. Then we refine this model using simulation based training in Section \ref{sec:pretrainalg}.
Next, in Section \ref{sec:quantawaretraining} we discuss quantization-aware training, which helps to ensure that the model remains accurate even when deployed in a quantized format for the purpose of reducing computational resources and speeding up the inference on resource-constrained programmable devices.
Then, in Section \ref{sec:hardwaretranslation} we discuss the process of hardware translation.
Next, in Section \ref{sec:arb} we discuss the adapted randomized benchmarking (ARB) algorithm for estimating gate fidelity for non-Clifford gates using measurements of the quantum device (required since direct observation of the wavefunction is not possible).
Following that, in Section \ref{sec:arbfinetune} we discuss a strategy for fine-tuning a bootstrapped model using ARB.
Finally, in Section \ref{sec:conclusion}, we conclude the paper.

\section{Methods}
\subsection{Data generation: creating quantum control samples with \texttt{Juqbox}}
\label{sec:datagen}

Quantum control sample generation via \texttt{Juqbox} \cite{petersson2022optimal} is a cornerstone of our methodology, as it defines the parameters within which we validate our hypotheses. Our objective was to produce samples comprising a single input angle in the range of $-\pi$ to $\pi$, paired with output pulse coefficients of a predefined size (20 in this case).
The qubit is controlled by a microwave control channel capacitively coupled to the qubit.
we use quadratic B-splines as the basis functions to decompose the envelop of the microwave control drive with the wave frequency set to be the qubit frequency. The first/last 10 pulse coefficients are the B-spline coefficients of the real/imaginary part of the control. In this study, we assume the qubit anharmonicity to be 200 MHz unless otherwise specified. We set the pulse duration time to be 125 ns and the maximum pulse amplitude to be 20 MHz for optimal control.
To implement the control pulse with specific hardware, the control pulse has to be converted to the voltage level of the control line. The conversion is typically determined by calibrating the quantum hardware \cite{Schuster2007,johnson2011controlling}. The calibration’s details are highly specific to the experimental setup and are outside the scope of this paper.

The generation of these samples was facilitated through a Jupyter Notebook\footnote{\url{https://jupyter.org}} script, which in turn invoked several \texttt{Julia}\footnote{\url{https://julialang.org}} scripts for configuration and generation using \texttt{Juqbox}.
Aiming for a high-fidelity threshold, we achieved a fidelity greater than 0.9999 (referred to as “four 9s” of Fidelity), signifying an excellent quantum state overlap. This level of precision, quantified by fidelity, highlights the data's precision and the system's effectiveness in maintaining the desired state.

The \texttt{Juqbox} mathematical model uses a seed value that alters the output pulse coefficients even when the input angle value remains unchanged. This creates a challenging dataset for model building due to large variations in outputs for neighboring input angles. To address this, we generate datasets using 100 seeds, each containing 4,096 input angles uniformly distributed from $-\pi$ to $\pi$, which helps reduce erratic variation as discussed in the next section. In addition to the general variation across different seeds, we observed that the pulse coefficients would invert at around -3.118 when initialized randomly.
We addressed this by including a fixed positive or negative baseline initial pulse based on the sign of the rotation angle, in addition to the random initialization in the \texttt{Julia} script responsible for generating these pulses by optimal pulse control. This fixed baseline serves as an educated guess to force the optimized pulses to carry the same sign as the baseline. Empirical analysis determined that approximately 4,096 samples within the $-\pi$ to $\pi$ range were ideal. Excess samples introduced unwanted noise, while too few samples reduced the model's ability to generalize.

\subsubsection{Data analysis: refining and optimizing quantum control samples}
\begin{figure}[t]
\centerline{\includegraphics[width=0.7\textwidth]{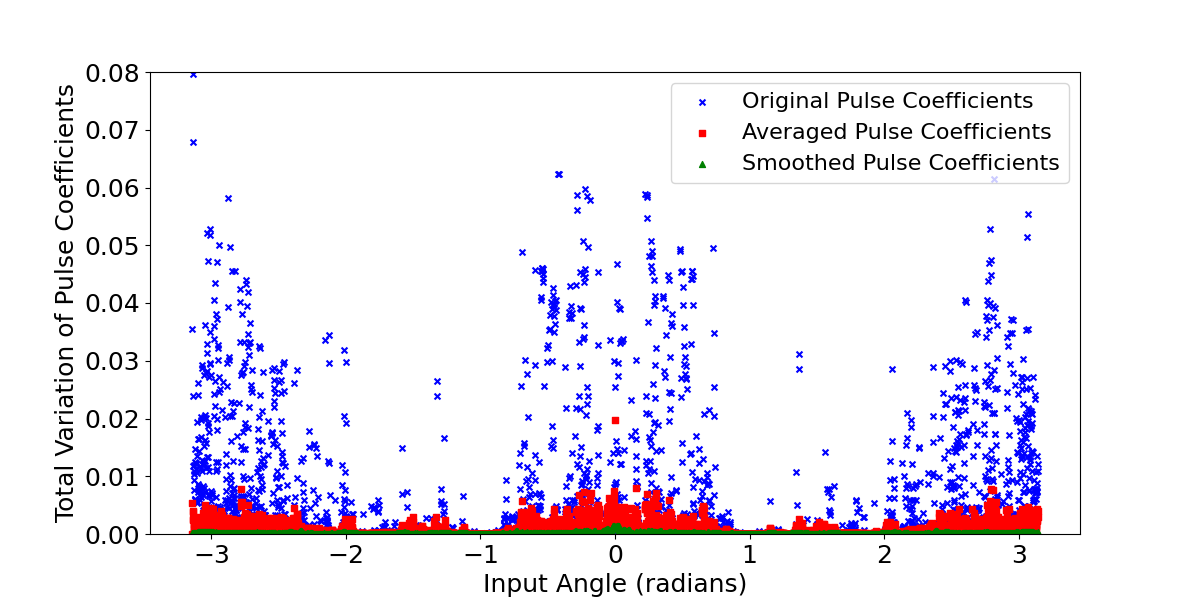}}
\caption{Total Variation of Pulse Parameters Over Angle. This overlay plot compares the original, averaged, and smoothed datasets, highlighting changes in parameter stability across different processing stages.}
\label{fig:total_variation}
\end{figure}

\begin{figure}[t]
\centerline{\includegraphics[width=0.95\textwidth]{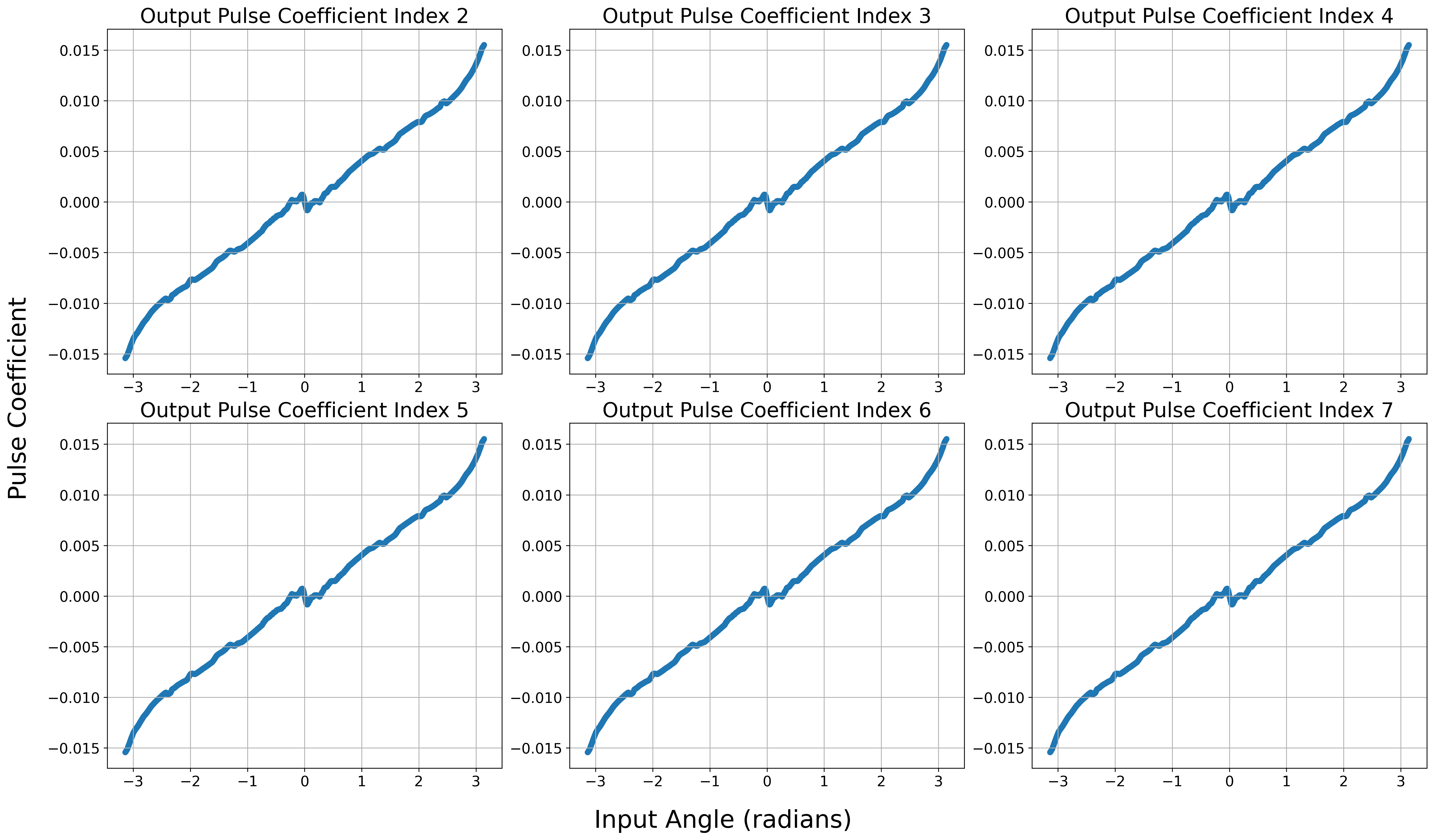}}
\caption{
Comparative subplots of output B-spline coefficients 2 to 7 out of 20 coefficients for X gates, demonstrating parallel trends. 
The six coefficients shown have similar values (y-axis) for a given X gate rotation angle (x-axis), and hence they can be replaced by the average value for data smoothing as explained in the main text.
}
\label{fig:param_trend}
\end{figure}

\begin{figure}[t]
\centerline{\includegraphics[width=0.7\textwidth]{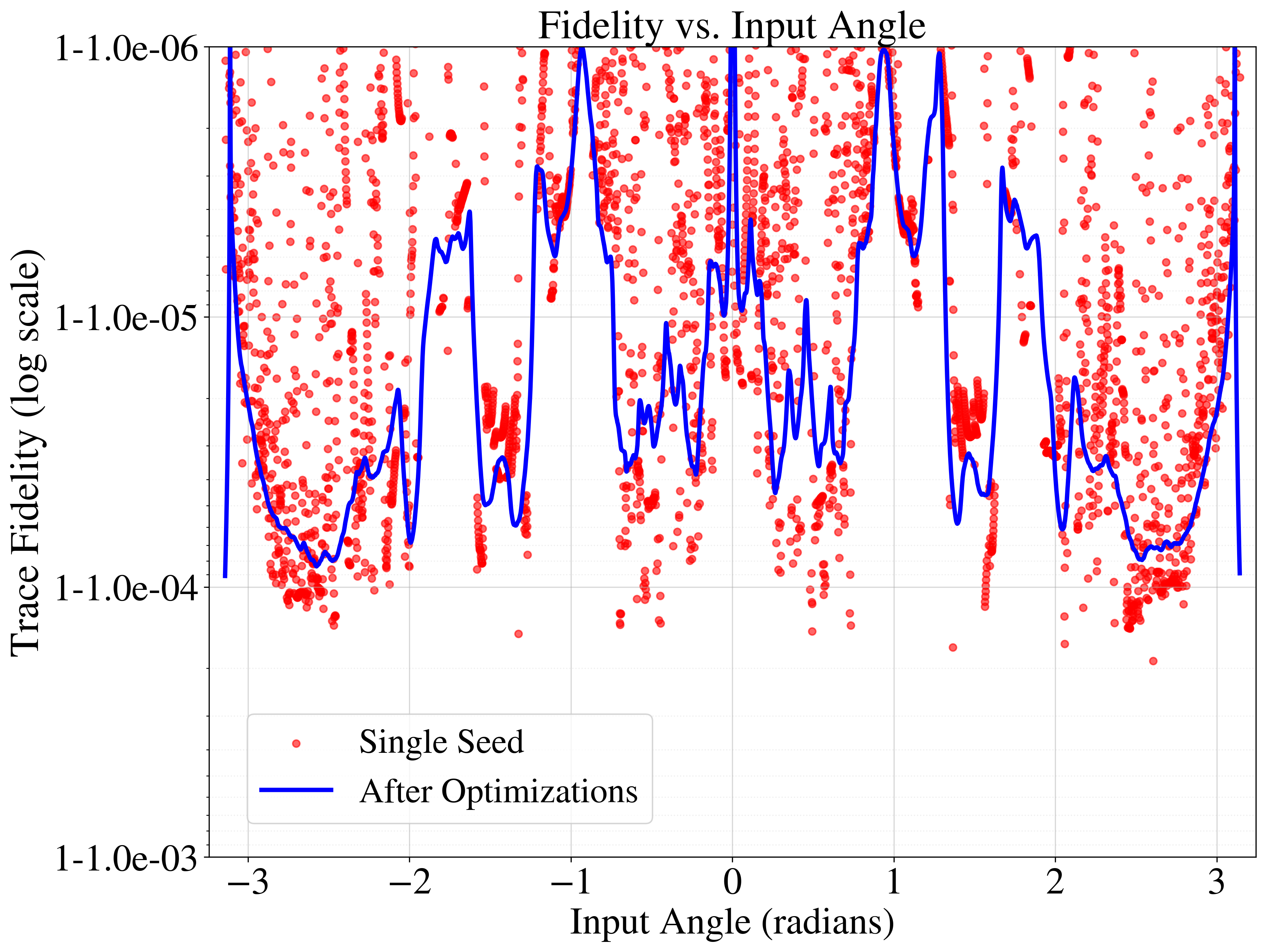}}
\caption{Trace Fidelity comparison between a single seed and after dataset optimizations over varying input angles.}
\label{fig:fidelity_compare}
\end{figure}

Refining and optimizing data samples is crucial for enhancing the performance and efficiency of model development. The process began with the organization and improvement of data generated from the 100 varied seeds, with each one producing slightly varied outputs. 

The uniformity of the data was improved by averaging the outcomes from all seeds, resulting in a single set of 4,096 samples. 
This averaging process is performed across the 100 seeds for each of the 4,096 input angles. Our objective is to mitigate the noise associated with any particular seed. For each input angle, the averaging can be expressed as:

\begin{equation}
A_i = \frac{s_{i1} + s_{i2} + s_{i3} + \cdots + s_{i100}}{100}
\end{equation}

\noindent
where $A_i$ is the averaged value for the $i$-th input angle, and $s_{ij}$ represents the value from the $j$-th seed for the $i$-th input angle. This process is repeated for all 4,096 input angles.
 
This step was critical to eliminate outliers and anomalies inherent in any individual seed. The data was then smoothed by using a sort of convolution that works by averaging groups of 50 close samples and using this average for the middle sample. This method helps the values change gradually instead of suddenly, leading to much less variation in the data, as shown in Fig. \ref{fig:total_variation}.
We further simplified our data for X gates by reducing the number of coefficients from 10 to 5. We observed that several coefficients exhibited minimal variation across different input angles, allowing us to replace them with their average values without significant loss of information. Figure \ref{fig:param_trend} illustrates this pattern for X gates.
Once the data was smoothed in this manner, it was divided into training, testing, and validation sets for use in subsequent stages. The script detailing these processes is available for reference.

Through these modifications, the dataset was condensed while maintaining four nines of fidelity. Despite a negligible decline in overall fidelity of the dataset ($< 0.00002$), the fidelity for some individual angle measurements actually improved, particularly in previously worst-case scenarios, as evidenced by Fig. \ref{fig:fidelity_compare}.

\subsection{Model training: optimizing for efficiency and fidelity}
\label{sec:pretrain}

\begin{figure}[t]
\centerline{\includegraphics[width=0.8\textwidth]{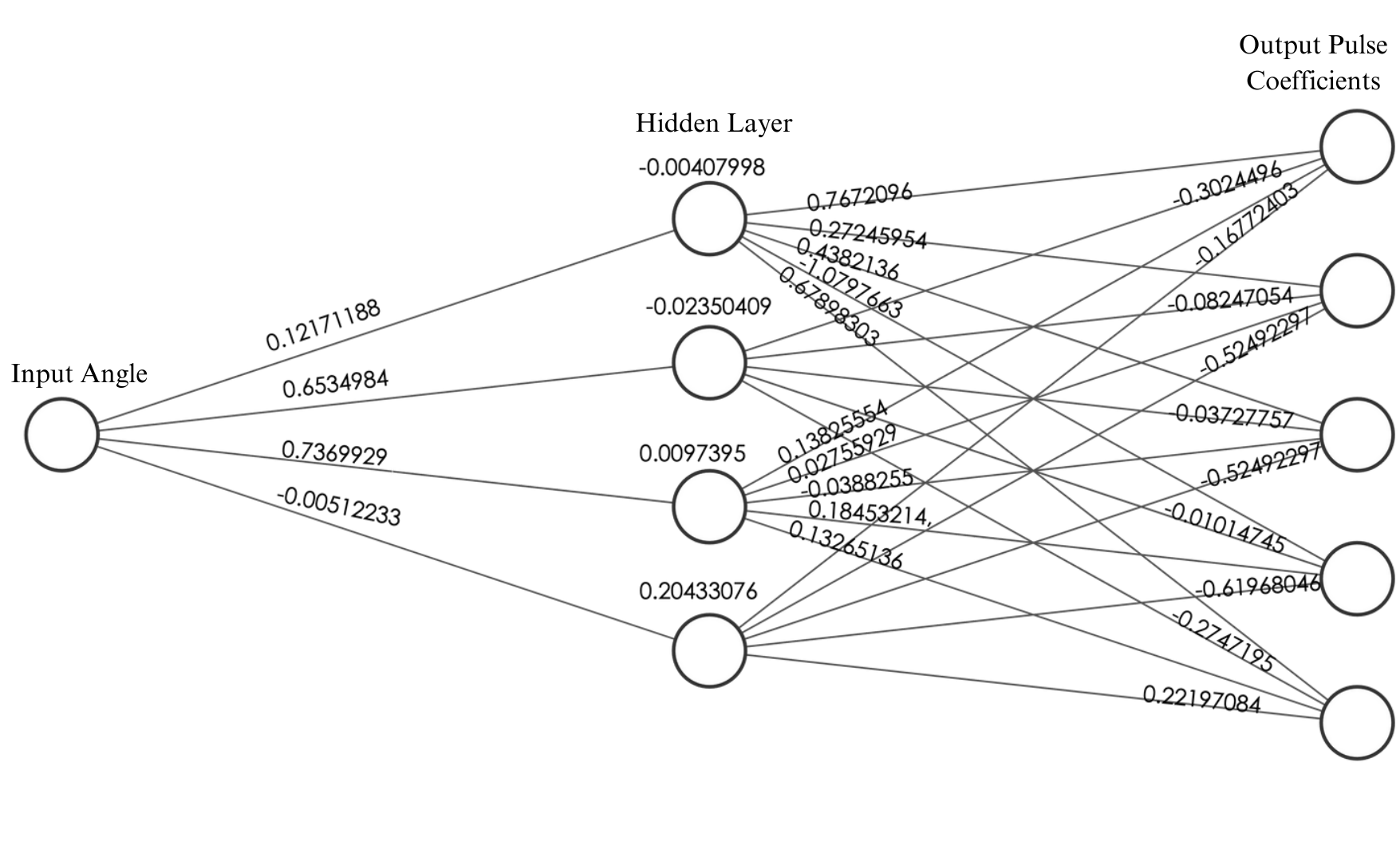}}
\caption{Illustration of the smallest Keras model configuration with 33 parameters achieving four nines of fidelity.}
\label{fig:model}
\end{figure}

For our model architecture, we opted for a multi-layer perceptron neural network due to the relatively straightforward nature of pulse coefficient prediction and its capability to efficiently translate to hardware. Our Keras-based neural network model's training process involves progressive stages. We begin with pre-training using `ground truth' values from quantum simulations, helping the model recognize similarities in pulse coefficients. Subsequently, we refine the model by focusing on quantum state output fidelity, using infidelity (1 - fidelity) as the loss function. This stage incorporates quantum simulation directly into training but is slower due to the non-analytical nature of the infidelity loss function. These steps are detailed in Section \ref{sec:pretrain}.

We then make training quantization-aware (Section \ref{sec:quantawaretraining}), translate the model for field programmable logic (Section \ref{sec:hardwaretranslation}), and fine-tune based on measurement-only information (Section \ref{sec:arbfinetune}).

Our focus is on creating a compact, efficient Keras model maintaining four nines of fidelity. This involves optimizing architecture, activation functions, learning rate, and loss function. The model is trained and evaluated using separate training and validation sets.

Pre-training involved 10,000 epochs with a 0.0001 learning rate, using MSE as the primary loss function. For fine-tuning, we employed an infidelity-based loss function, introducing controlled perturbations to the model's weights and recalculating infidelity. This process, detailed in Section \ref{sec:pretrainalg}, significantly improved performance but is considerably slower than MSE training.

Through these optimizations, we reduced model complexity from over 2000 to just 33 adjustable parameters. Fig. \ref{fig:model} illustrates the final model. This neural network model has a single input with a single hidden layer of size 4 and an output layer of size 5 which is reduced from the 20 pulse coefficients as explained earlier in Section \ref{sec:datagen}.
  
\subsubsection{Simulation-based training algorithm}
\label{sec:pretrainalg}

The algorithm is built on a set of functions that are available in our software package~\cite{github}.
See Algorithm \ref{alg:simbasedtrain}.
\begin{itemize}
     \item \textbf{Function \texttt{infidelity\_loss\_parallelized(x, y\_preds, y\_orig)}:}
    \begin{itemize}
        \item \emph{Description}: Calls an external quantum simulation to compute infidelities in parallel.
        \item \emph{Input}: $x$ - input data, $y\_preds$ - predicted values, $y\_orig$ - original values.
        \item \emph{Output}: $infidelities\_y\_preds$ - infidelities of $y\_preds$, $infidelities\_y\_orig$ - infidelities of $y\_orig$.
    \end{itemize}
    
    \item \textbf{Function \texttt{infid\_grad(x, model, epsilon)}:}
    \begin{itemize}
        \item \emph{Description}: Applies a small epsilon perturbation to each weight/bias of the model and computes the gradients. The gradients are obtained by calculating the infidelities using infidelity\_loss\_parallelized, computing the difference before and after the perturbation, and dividing that by the perturbation epsilon.
        \item \emph{Input}: $x$ - input data, $model$ - trained model, $epsilon$ - small perturbation value.
        \item \emph{Output}: $gradients$ - computed gradients, $infidelity\_original$ - original infidelity.
    \end{itemize}
    
    \item \textbf{Function \texttt{train\_step(x\_batch, y\_batch)}:}
    \begin{itemize}
        \item \emph{Description}: Computes gradients and loss for the batch using infid\_grad, updates model trainable variables using optimizer, updates loss metric to track loss.
        \item \emph{Input}: $x\_batch$ - batch of input data, $y\_batch$ - batch of output data.
    \end{itemize}
    
    
\end{itemize}

\begin{algorithm}
\caption{Simulation-based training}
\label{alg:simbasedtrain}
\begin{algorithmic}[1]
\State Load the best pre-trained model based on an MSE loss over pulse coefficients.
\State Initialize the optimizer and loss metric.
\State Set up \texttt{ModelCheckpoint} callback with the specified coefficients.
\For{epoch in training epochs}
    \State Reset loss metric.
    \For{batch in training set}
        \State Perform \texttt{train\_step} with \texttt{x\_batch} and \texttt{y\_batch}.
    \EndFor
    \State Update model checkpoint.
\EndFor
\end{algorithmic}
\end{algorithm}

\subsection{Quantization aware training}
\label{sec:quantawaretraining}

Quantization-aware training (QAT) is critical in optimizing ML models for efficient deployment on hardware accelerators.
This is particularly true for field-programmable logic, which is increasingly used in edge computing due to its reconfigurability, energy efficiency, and ability to perform parallel computations. However, constrained resources, such as limited memory and computational elements, necessitate deploying carefully tailored models. 
Thus, the primary motivation behind QAT is to reduce the precision of the weights and activations in neural networks from floating-point to fixed-point representations, thereby decreasing the model size and computational complexity. By training neural networks to be aware of quantization effects, it is possible to significantly mitigate the degradation in performance typically associated with more traditional techniques like post-training quantization.

In our work, we have adopted \texttt{QKeras}~\cite{coelho2021automatic}, an extension of the popular \texttt{Keras} library that mimics the behavior of fixed-point arithmetic as part of the training process. \texttt{QKeras} provides quantized versions of standard \texttt{Keras} layers (e.g., \texttt{QDense}, \texttt{QConv2D}) where the designer can specify the bit width for weights, biases, and activations directly in the layer definitions. During training, \texttt{QKeras} simulates the quantization process: the forward pass computes the layer outputs using quantized weights and activations, simulating the effects of fixed-point quantization. However, for the backward pass and weight updates, floating-point precision is typically used to maintain training stability and performance.
%
%
%
%
Among the \texttt{QKeras} customizable parameters, we experimented with:
\begin{itemize}
\item \textbf{bits} to select the total fixed-point bit width ($W$) allocated for a layer. Our experiments indicate that diminishing this value to as low as 16 bits preserves our model fidelity.
\item \textbf{integer} to select the number of bits ($I$) allocated to the integer portion of the fixed-point representation. Maintaining a minimum of 5 bits for the integer part and 11 bits ($W-I$) for the decimal part did not significantly diminish our model fidelity.
\item \textbf{alpha} facilitates the simulation of Leaky ReLU functions. A Leaky ReLU introduces a nonzero gradient $alpha$ for negative inputs. Introducing this slight slope for negative values helps keep neurons ``alive'' by ensuring they can still learn during the backpropagation process even when their inputs are negative. In our initial experiments, we adopted a value of 1.
\item \textbf{qnoise\_factor} determines the extent to which quantization noise is added to the weights and activations during the forward and backward passes of model training. The network learns to cope with the noise, leading to potentially better generalization and accuracy in the quantized model. As the value of noise increases, more quantization noise is added, simulating a higher degree of quantization effect. In our current study, we set this parameter to 1.
\end{itemize}

Finally, it is worth noting that we did not retrain the model from scratch in \texttt{QKeras}. Instead, we transferred the weights from the previously trained \texttt{Keras} model to the quantized \texttt{QKeras} model and then additionally ran quantization-aware training. This method ensured the model's fidelity while transitioning to a quantized representation.

\subsection{Hardware translation: from quantized model to FPGA deployment}
\label{sec:hardwaretranslation}

We adopted \texttt{hls4ml}, a Python open-source framework \cite{vloncar_2021_5680908,Duarte:2018ite}, to co-design and translate our ML models into a hardware implementation while studying model accuracy, resource utilization, and inference latency.
The \texttt{hls4ml} workflow begins with a floating-point model from a conventional ML framework, such as \texttt{TensorFlow} or \texttt{PyTorch}, or a quantized model from \texttt{QKeras}.
Then, it translates the model into a C++ specification for the high-level synthesis (HLS) flow. HLS generates a hardware description at the register-transfer level for a more traditional synthesis and implementation flow targeting programmable logic as deployment hardware. 
Designers can leverage \texttt{hls4ml} to make quick design explorations by configuring the hardware implementation parallelism \cite{fahim2021hls4ml} and, thanks to the integration with \texttt{QKeras}, by also evaluating the impact of low-bit precision on model performance before finalizing the hardware implementation.


%
%
%

We iteratively translated the \texttt{QKeras} model into an HLS/C++ specification and hardware implementation to evaluate the resource utilization and model fidelity.
In particular, we tuned the reuse factor parameter in \texttt{hls4ml}, which impacts parallelism, resource utilization, and performance. In our experience, a reuse factor of 20 has shown a good compromise in resource utilization.
Moreover, we adjusted the accumulators' bit accuracy for each layer in fixed-point arithmetic. It is worth noting that \texttt{QKeras} does not provide the fine-tuning necessary for optimal hardware implementation. We observed a minimum bit width of around 22 to avoid performance degradation.
At this point, \texttt{hls4ml} automatically translates the ML model into an HLS/C++ specification that can be simulated for fidelity assessment. 
This verification step is crucial for confirming that the translation has been successful and that the model specification is ready for the hardware implementation.

We leveraged graphical tools provided by the \texttt{hls4ml} framework to ensure translation correctness. For example, we use identity-line plots to compare the layer outputs. In such a plot, the diagonal line (the line of identity) represents perfect agreement between \texttt{QKeras} and HLS/C++ implementations.
Points sitting precisely on the diagonal indicate that for every input of the layers, both implementations produce the same outputs. The goal is for all points to lie as close to the diagonal as possible, indicating that the two-layer implementations produce nearly identical results. Deviations from the diagonal suggest discrepancies between the outputs of the two systems and should be carefully analyzed.

We ran HLS and implementation targeting FlexLogix eFPGA, a reconfigurable fabric that offers efficient and flexible hardware acceleration solutions~\cite{FlexLogix}. The results of our implementation showed a resource utilization of 693 LUTs, 709 FFs, and 2 DSPs, with a latency of 420ns. The resource utilization is minimal even for the smallest configuration of FlexLogix eFPGAs.

%
%

\subsection{Adapted Randomized Benchmarking (ARB)}
\label{sec:arb}

The standard approach for estimating quantum gate fidelity in the literature is Randomized Benchmarking (RB), which applies to Clifford gates.
Here we provide an adapted algorithm for non-Clifford gates, updating the ideas behind RB, and provide a method for computing confidence intervals for the fidelity.

Assuming that only $R_x(\theta)$ gates are to be tested, let $\mathcal{G}$ be the set of angles that we want to test and $|\mathcal{G}|$ its cardinality.
For each angle $\theta_i$, the corresponding imperfect gate is $\hat{R}_x(\theta_i)$. 
Before starting the algorithm, there are two required steps:
\begin{itemize}
    \item Preliminary step 1: Choose a set of sequence lengths $M=\{m_1, \dots, m_M\}$.
    Each sequence length defines the number of gates that are consecutively applied to the initial state, where the last gate will be the inverse of the sum of the previous gates.
    The denser $M$ is the better the estimation will be. 
    We empirically observed $2$ to $100$ is a reasonable range.
    \item Preliminary step 2: Choose $K$ random sequences.
    Each $k^\text{th}$ sequence will be $m$ gates long, depending on the current sequence length $m$.
    The first $m-1$ gates are uniformly sampled (possibly with repetition) from the set to be tested.
    For each new sequence, a new sampling is performed, resulting in (typically) unique sequences.
    We empirically observed a $K$ on the order of hundreds to be performant.
\end{itemize}

With the preliminaries complete the Adapted Randomized Benchmarking algorithm for non-Clifford gates is Algorithm \ref{alg:arb}.
Although a formal relation between gate fidelity and the metric estimated by ARB is not established, we will show numerically that the estimation successfully characterizes the error in rotation angles and can be used as the cost function to train a neural network to correct the error.

\begin{algorithm}
\caption{Adapted Randomized Benchmarking for non-Clifford gates}
\label{alg:arb}
\begin{algorithmic}[1]
\For{$m \in M$}
    \For{$k \in {1 \dots K}$}
        \State Uniformly sample a random sequence of numbers $s_k = s_{1_k} s_{2_k} \dots s_{{m-1}_k}$, where $s_{i_k} \in \{1,\dots,|\mathcal{G}|\}$.
        Each $s_{i_k}$ corresponds to a gate $\hat{R}_x(\theta_{s_{i_k}})$.
        \State Prepare the initial state $\ket{0}$.
        \State Apply the sequence of gates $\hat{R}_x(\theta_{s_{1_k}})\dots\hat{R}_x(\theta_{s_{m-1_{k}}})\hat{R}_x(-\sum_{i=1}^{m-1} \theta_{s_{i_k}})$.
        \State Perform $N$ repetitions and measurements in the Pauli-Z basis. 
        The probability of outcome $0$ is estimated as $$\hat{p}_{k,m} = \frac{\#zeros}{N},$$ with standard error from the Binomial distribution, $$SE_{\hat{p}_k,m}=\frac{\sqrt{\hat{p}_{k,m}(1-\hat{p}_{k,m}})}{N}$$.
    \EndFor
    \State Calculate the average of $\hat{p}_m$, $\text{avg}_{\hat{p},m}$, over the $K$ sequences. Its standard error will be $$SE_{{\text{avg}}_{\hat{p},m}} =\frac{\sqrt{\sum{SE^2_{\hat{p}_{k,m}}}}}{K}.$$
    \State The current estimated fidelity with $m$ gates is therefore $\mathcal{F}_m = \text{avg}_{\hat{p},m}$, with standard error $err_m =  SE_{{\text{avg}}_{\hat{p},m}}$.
\EndFor
\State After calculating the quantities for each $m$, fit $\bar{\mathcal{F}}_m = A+Bf^m$.
Here, $A, B, f$ are the coefficients to be found, $m$ is the independent variable and $\bar{\mathcal{F}}_m$ the dependent variable.
The $err_m$'s are used as uncertainties to obtain the resulting uncertainties on $A, B$ and $f$.
The bounds should be $0\leq A \leq 1$, $0 \leq B \leq 1$, $0 \leq f \leq 1$.
\State We now have an estimate $\hat{f}$ for the single-gate fidelity from the fit, along with a covariance matrix\footnote{For example, this is provided automatically when using \texttt{scipy.optimize.curve\_fit} \cite{scipy.optimize.curvefit}.} for the three parameters, $\Sigma \in \mathbb{R}^{3,3}$. 
The third diagonal entry (if $f$ was used as third parameter for the fitting) will contain $SE^2_{\hat{f}}$.
\State Finally, perform a 2-tailed Student's t-test, where the number of degrees of freedom is equal to $|M|-3$ (3 is the number of parameters), and setting $\alpha = 0.05$ to obtain a $95\%$ confidence interval for $f$, corresponding to $$\left[\hat{f}-t_\alpha\sqrt{\Sigma(3,3)}, \hat{f}+t_\alpha\sqrt{\Sigma(3,3)}\right]$$.
\end{algorithmic}
\end{algorithm}

\subsubsection{Testing ARB with direct unitaries}

As an initial demonstration, consider a set of perturbed gates without an underlying physical simulation.
Here, the interval $\left[-\pi,\pi\right]$ is divided in 1,000 angles and each angle $\theta_i$ is perturbed by adding noise term drawn from a Gaussian distribution, $\hat{\theta_i}=\theta_i+\mathcal{N}(0,\sigma)$.
We perform ARB by doing a sequence of rotations using these perturbed angles, but we modify the procedure and set the final rotation as the inverse formed from the sum of the exact angles instead the noisy angles.
While ARB shouldn't use exact gates, here it is necessary because if we use the sum of the perturbed angles for the inverse, we would obtain exactly the initial state.
An alternative would have been to use the sum of the same angles and add another Gaussian noise perturbation.

\begin{figure}[t]
\centerline{\includegraphics[width=1\textwidth]{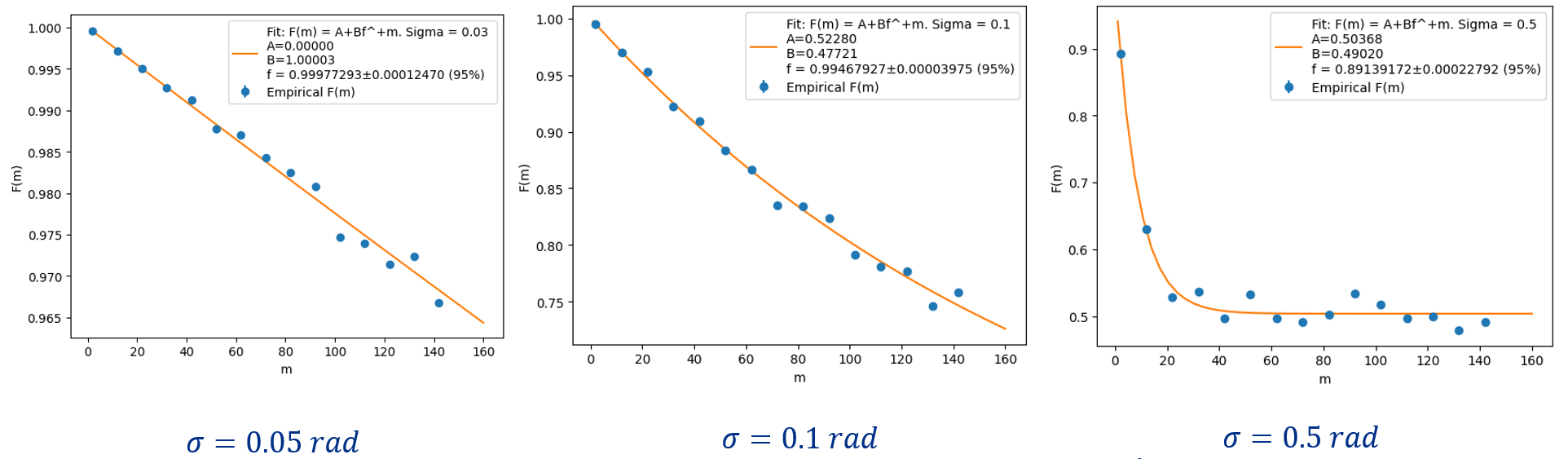}}
\caption{The first three experiments with artificially perturbed gates (K=500).}
\label{fig:K500}
\end{figure}

\begin{figure}[t]
\centerline{\includegraphics[width=1\textwidth]{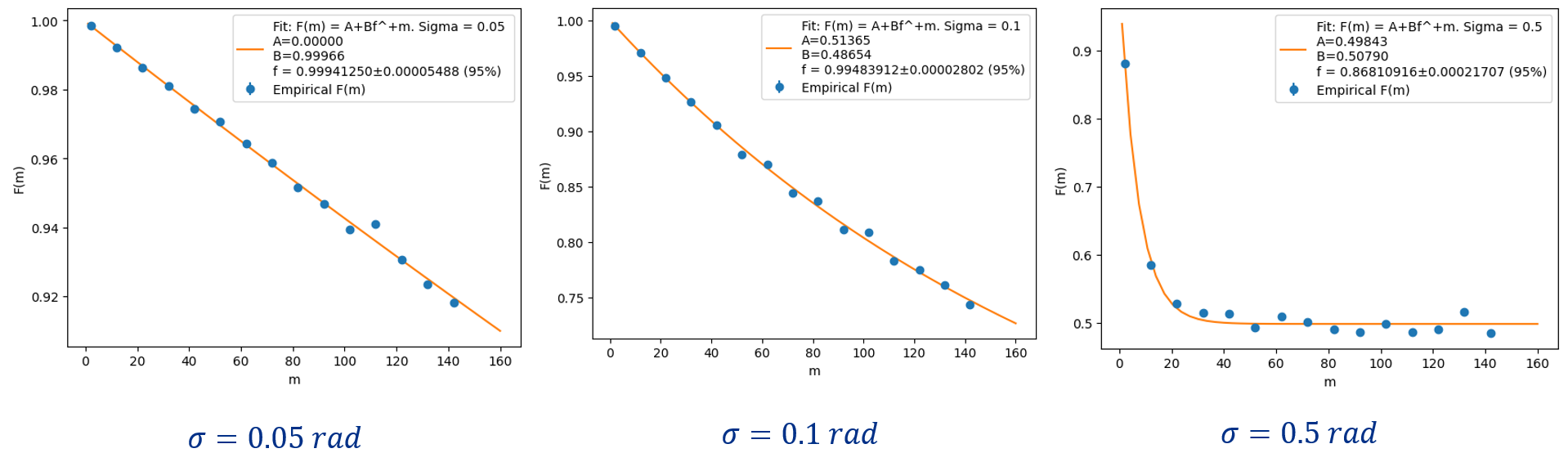}}
\caption{The last three experiments with artificially perturbed gates (K=1,000).}
\label{fig:K1000}
\end{figure}

We perform six experiments, varying $K$ and $\sigma$.
In all of them we set $N=1,000$ and $M=\texttt{range}(2,150,10)$.
The first three have $K=500$, and $\sigma$ equal to $0.05$ rad, then $0.1$ rad, and then finally $0.5$ rad. 
The results are shown in Fig. \ref{fig:K500}.
The last three have $K=1,000$, and the same $\sigma$ values. 
These results are shown in Fig. \ref{fig:K1000}.

\subsubsection{Using ARB for testing a pre-trained neural network on different physical conditions}

We initially trained a neural network using a quantum simulation of a qubit with 0 guard levels instead of 1 guard level.
The network was trained using the Mean-Squared Error (MSE) between the pulse coefficients generated by Juqbox for the angles in the training set and the corresponding pulse coefficients inferred by the NN as a loss function. 
We used ARB to measure the fidelity using a simulation with zero guard levels and identical anharmonicity, finding a fidelity that remained at the level of four nines.

The actual physical system has an infinite number of guard levels, and the occupancy probability falls as the level increases.
Additionally, temperature cycles such as heating up the system and then cooling it down again impact some physical parameters, such as the qubit anharmonicity.
Therefore, we expect the performance of the NN to decrease over time, because it was trained on pulses that \texttt{Juqbox} generated utilizing different physical parameters.

We examined this network using ARB to measure the fidelity of the pulses generated by the pre-trained neural network, first setting the anharmonicity to $200$ MHz, then multiplying it by 10.
In both cases, we added one guard level to the simulation.
The primary purpose of this change is to simulate a mis-modeling of the physics parameters that describe the device.
This allows for some estimation of the impact of imperfect device simulation on NN training, for example, in the case of using simulation to bootstrap a model for fine-tuning on hardware.
It also provides a proxy for understanding the impact of drifts in the device noise characteristics, which is an important issue for keeping a NN model well-tuned.

\begin{figure}[t]
\centerline{\includegraphics[width=0.7\textwidth]{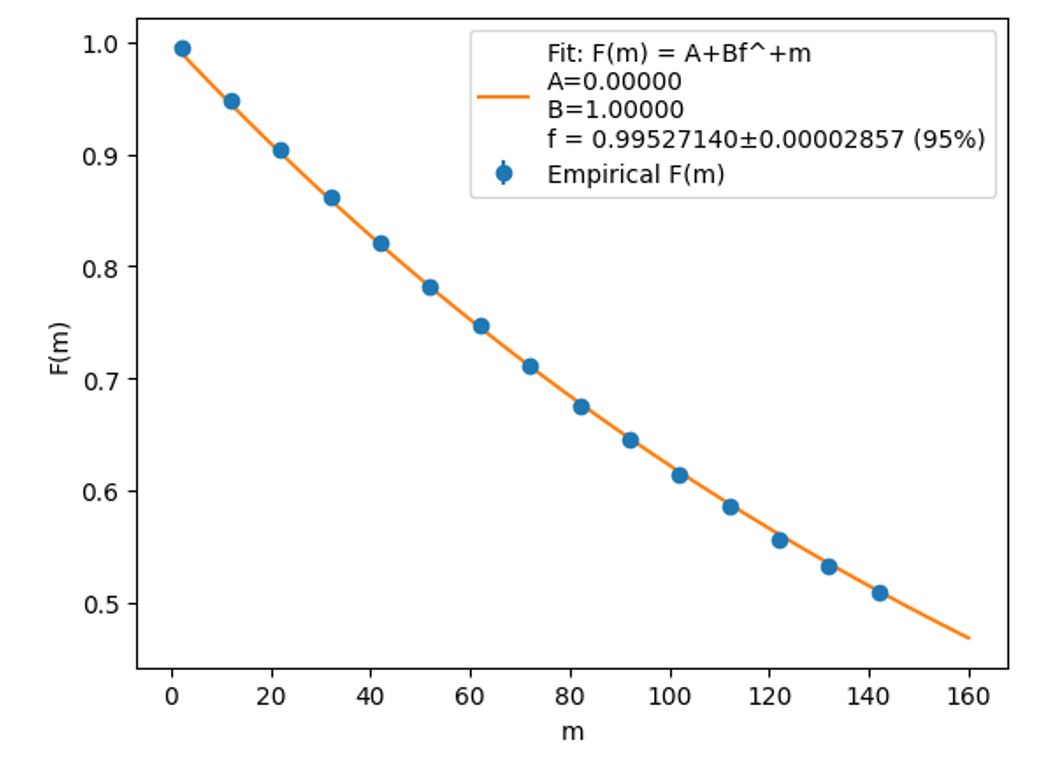}}
\caption{Fidelity estimated with ARB when the anharmonicity is unchanged.}
\label{fig:anharm_1}
\end{figure}

\begin{figure}[t]
\centerline{\includegraphics[width=0.7\textwidth]{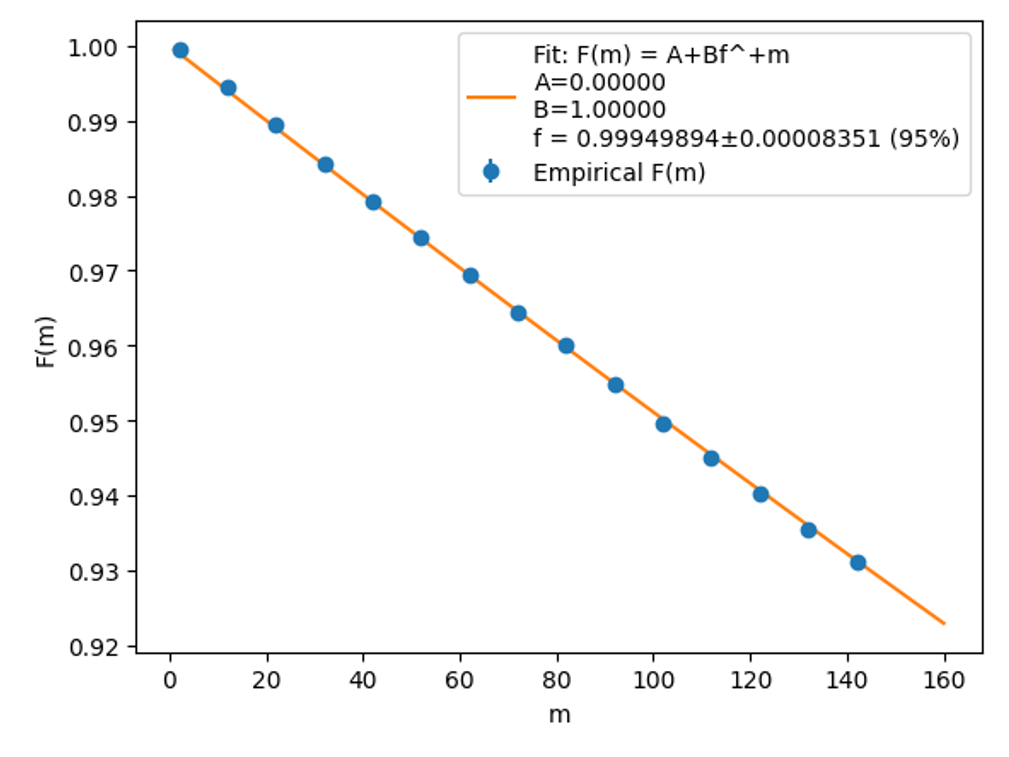}}
\caption{Fidelity estimated with ARB when the anharmonicity is x10 the one used during training.}
\label{fig:anharm_10}
\end{figure}

As shown in Fig. \ref{fig:anharm_1}, introducing one guard level causes the fidelity decrease by two orders of magnitude.
In Fig. \ref{fig:anharm_10}, however, it is evident that multiplying the anharmonicity by 10 has the effect of compensating for this by moving the guard level further from the two essential levels.

\section{Results}

\subsection{Using ARB for fine-tuning}
\label{sec:arbfinetune}

Our strategy for adapting ARB to a realistic scenario is to take the pre-trained network (as described in Section \ref{sec:pretrain}) and to fine-tune it to a different configuration, using ARB instead of the MSE with ``ideal'' pulse coefficients for the NN loss.

In particular, a simulated scenario with G=1 guarded levels is considered, with a new anharmonicity equal to $2$ GHz, which is 10 times the one that the network was trained on.
Notice that this higher anharmonicity compensates for the new guarded level, as it provides a higher frequency separation of the two essential levels $\ket{0}, \ket{1}$ from the guarded level $\ket{2}$. 

\subsection{Code structure}
The NN model was trained in \texttt{Python}\footnote{\url{https://www.python.org}} using \texttt{TensorFlow} \cite{tensorflow2015-whitepaper}.
We further use \texttt{Python} to do the fine-tuning and the inference.
At each training epoch, we consider a set of angles from $-\pi$ to $\pi$ and infer a list of $20$ pulse coefficients for each of these angles. 
The 20 coefficients are the B-spline coefficients of the control pulse described in section \ref{sec:datagen}.
Then, for each angle, its corresponding list of coefficients is used to make the system evolve according to the corresponding pulse and a unitary corresponding to that pulse is obtained.
Doing this for all the angles, a set of unitaries is obtained corresponding to the gates that the network inferred for all the angles.
At this point, ARB is run to measure the fidelity of these gates to the theoretical ones, and the weights of the network are updated accordingly via gradient descent.

A technical complication arises from the fact that Juqbox is implemented in \texttt{Julia}, so we transfer data between applications using pipes, which are a type of Inter-Process Communication (IPC), exchanging data in JSON\footnote{\url{https://www.json.org/json-en.html}} format.

Another technical consideration is that the unitaries obtained with \texttt{Juqbox} result from a numerical integration of differential equations, and due to limited precision may not be unitary.
Therefore, as an additional step we renormalize them using their 2-norm.

\subsubsection{Optimization algorithm for training}

Our ultimate goal is to deploy this network on an embedded device, so we invested substantial effort in minimizing the footprint of the NN.
The ``small'' version of the network has 8 dense layers (fully connected layers with ReLu activation function), resulting in $760$ parameters (weights and offsets). 
Because our loss is calculated via IPC program calls, \texttt{TensorFlow}\texttt{autograd} cannot be used, and gradients must be computed by hand.
Exact gradient computation would require calculation of the loss varying each parameter.
By doing this for all the parameters we would estimate the gradient.
However, this results in the calculation of $760 * 2$ losses for each epoch, where each loss calculation involves a number of simulations equal to the number of angles used in training, which can be in the order of thousands.
While accurate this procedure is unacceptably slow.

\begin{figure}[t]
\centerline{\includegraphics[width=0.8\textwidth]{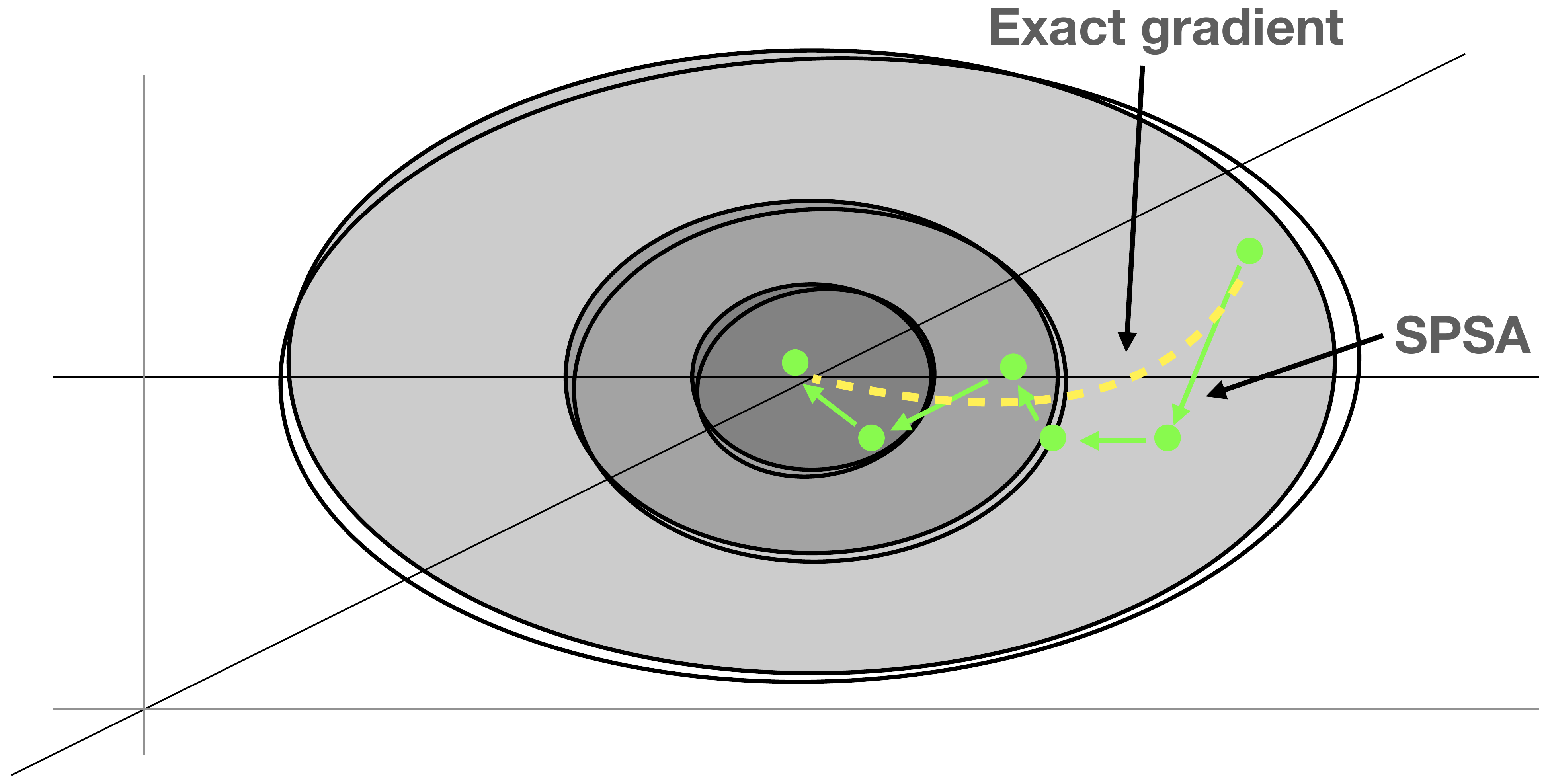}}
\caption{An illustration showing how SPSA works compared with standard gradient descent (figure inspired by \cite{QiskitTextbook}).}
\label{fig:spsa}
\end{figure}

Another method, proven to statistically converge to the same solution, is Simulatenous Perturbation Stochastic Approximation (SPSA) \cite{119632} (Fig.~\ref{fig:spsa}). 
Here, all network parameters are perturbed at once, adding the same $\epsilon$ to all of them but with random sign. 
Specifically, considering a function $f(\vec{\theta})$ of which we want to estimate the gradient, we perform
\begin{equation}
    \nabla f(\vec{\theta}) = \begin{bmatrix}
        \frac{\partial f}{\partial \theta_1} \\
        \dots \\
        \frac{\partial f}{\partial \theta_n} \\
    \end{bmatrix} \approx \frac{f(\vec{\theta}+\epsilon\vec{\Delta})-f(\vec{\theta})}{\epsilon} \vec{\Delta}^{-1}
\end{equation}
where $\vec{\Delta} \in \{+1,-1\}^n$ and, in this case, $\vec{\Delta}^{-1}=\vec{\Delta}$ being its element-wise reciprocal.
The advantage of SPSA is that it only requires $2$ evaluations per epoch, one with the perturbed parameters and one with the unvaried ones, independent of the network size.

During training, there are actually two hyperparameters that determine the ``aggressiveness'' of gradient descent.
One is the $\epsilon$ by which the gradient is estimated, and the other is the learning rate $\alpha$, where $\vec{\theta}_{k+1} = \vec{\theta}_{k}-\alpha{\hat{\nabla}{f}}(\vec{\theta}_k)$.

\subsubsection{Training set construction}

There are multiple ways to construct the training dataset. 
Network generalization is generally enhanced by a larger number of angles used for training.
We empirically observed $500$ angles to be a good starting point.
Our first experiment used a training set with $500$ uniformly sampled angles, with a similar construction for the validation dataset with $100$ angles.
Note that the sampling procedure generates spacing that is generally non-uniform.
As a form of regularization, then, the training set was resampled at each epoch.

Then, for a second experiment, we trained a version of the network with fixed intervals between angles without re-sampling.
The training set was further divided into 10 batches.
The validation set was again uniformly sampled.
Finally, $\alpha = \epsilon = 10^{-6}$ were used for both experiments.

\subsubsection{Results}

\begin{figure}[t]
\centerline{\includegraphics[width=0.7\textwidth]{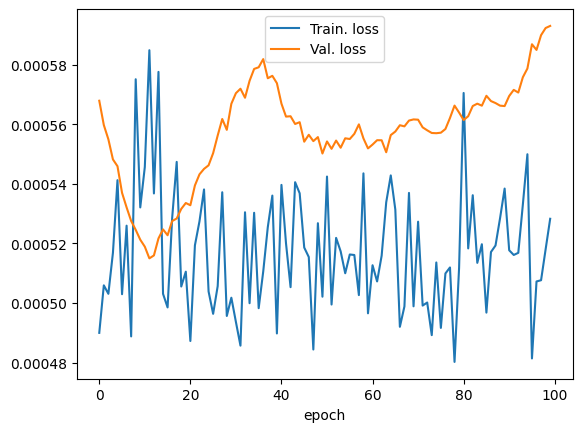}}
\caption{Loss function during training for the first experiment. Blue = training loss, orange = validation loss}
\label{fig:fine_tuning_1}
\end{figure}

\begin{figure}[t]
\centerline{\includegraphics[width=0.7\textwidth]{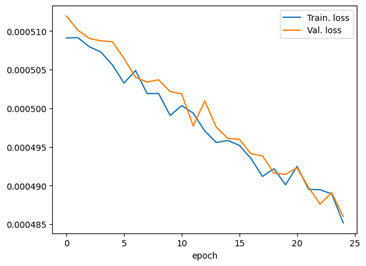}}
\caption{Loss function during training for the second experiment.}
\label{fig:fine_tuning_2}
\end{figure}

The results of the first experiment are shown in Fig. \ref{fig:fine_tuning_1}.
We found re-sampling to construct the training set at each epoch made the learning unstable for this problem.
In Fig. \ref{fig:fine_tuning_2}, we show the results of the second approach and find better results.

\section{Discussion}
\label{sec:conclusion}

From the results, we see that many epochs are needed to achieve a consistent improvement of accuracy. 
It is likely that, if the training is uncontrolled, the loss will at some point stabilize or start increasing again.
For this reason, an early stopping mechanism is necessary, and some kind of regularization could be useful to avoid overfitting.
The loss function is highly non-convex so it's very hard to achieve its global minimum, and a hyperparameter tuning process can help reaching the best possible local minima.
A larger neural network may offer better performance, but at the cost of exceeding resources on an embedded platform.


Another potential improvement would be to modify the architecture of the network to include the qubit anharmonicity among the inputs, together with the rotation angle.
This may reduce the need to retrain the network each time the anharmonicity changes, if it could reliably be trained once considering a set of possible anharmonicities.

In conclusion, Adapted Randomized Benchmarking is a good strategy for optimizing neural networks to work with varying physical conditions, making it possible to potentially achieve a high fidelity arbitrary rotation gate on hardware platforms.
Because this work is conducted on a simulation of a quantum device, we cannot guarantee the observed results will translate to hardware.
However, we provide a complete workflow for real quantum hardware.

While we have not demonstrated the technique on a real hardware platform, by utilizing varying simulation configurations without utilizing the underlying knowledge of those configurations for tuning, our results suggest these techniques will translate to hardware.
However, these techniques are not likely to be shot-efficient for superconducting transmons if noise drifts on a platform are significant, as re-training the NN will be time-consuming relative to the strategy of calibrating fixed angle gates and composing arbitrary rotations by utilizing virtual Z's. 
This technique will be more useful for platforms that do not have the benefit of a ``free'' and very high-fidelity rotation gate.

\bmhead{Acknowledgements}
\label{sec:ack}

M.R. was partially supported for this work by the Summer Students Italian Program at the Fermi National Accelerator Laboratory.
He would also like to thank Bartolomeo Montrucchio for extensive support in the role of PhD advisor.
M.R., G.P., and A.C.Y.L. were partially supported for this work by the DOE/HEP QuantISED program grant ``HEP Machine Learning and Optimization Go Quantum,'' identification number 0000240323.

G.D.G. is supported by Fermi Research Alliance, LLC under Contract No. DE-AC02-07CH11359 with the Department of Energy (DOE), Office of Science, Office of High Energy Physics.

This document was prepared using the resources of the Fermi National Accelerator Laboratory (Fermilab), a U.S. Department of Energy (DOE), Office of Science, HEP User Facility. 
Fermilab is managed by Fermi Research Alliance, LLC (FRA), acting under Contract No. DE-AC02-07CH11359.

Fermilab report number: FERMILAB-PUB-24-0127-ETD-SQMS.

\section*{Declarations}

\subsection*{Competing interests}
The authors declare no competing interests.

\bibliography{references.bib}


\end{document}